\documentstyle{mn}

\def\spose#1{\hbox to 0pt{#1\hss}} 
\def\simlt{\mathrel{\spose{\lower 3pt\hbox{$\mathchar"218$}}
     \raise 2.0pt\hbox{$\mathchar"13C$}}}
\def\simgt{\mathrel{\spose{\lower 3pt\hbox{$\mathchar"218$}}
     \raise 2.0pt\hbox{$\mathchar"13E$}}}

\begin{document}

\title[Optical photometry of CAL~87]
{Optical photometry of the eclipsing Large Magellanic Cloud 
supersoft source CAL~87}

\author[C. Alcock et al.]
{C. Alcock$^{1,2}$, R.A. Allsman$^{3}$, D. Alves$^{1}$, T.S. Axelrod$^{1,4}$,
D.P. Bennett$^{1,2}$,  \and P. A. Charles$^{5}$, 
K.H. Cook$^{1,2}$, D.\ O'Donoghue$^{6}$, 
K.C. Freeman$^{4}$, K. Griest$^{2,7}$, \and J. Guern$^{2,7}$,
M.J. Lehner$^{2,7}$, M. Livio$^{8}$, 
S.L. Marshall$^{2,9}$, D. Minniti$^{1}$, \and 
B.A. Peterson$^{4}$, M.R. Pratt$^{2,9}$,  
P.J. Quinn$^{4}$, A.W. Rodgers$^{4}$, 
K.A. Southwell$^{5}$, \and C.W. Stubbs$^{2,9,10}$,  
W. Sutherland$^{5}$ and D.L. Welch$^{11}$\\
$^1$ Lawrence Livermore National Laboratory, Livermore, CA 94550, USA\\
$^2$ Center for Particle Astrophysics, University of California, Berkeley, 
CA 94720, USA\\
$^3$ Supercomputing Facility, Australian National University, Canberra, A.C.T. 
0200, Australia\\
$^4$ Mt.  Stromlo and Siding Spring Observatories, 
Australian National University, Weston, A.C.T. 2611, Australia\\
$^5$ University of Oxford, Department of Astrophysics, Nuclear \& Astrophysics
Laboratory, Keble Road Oxford, OX1 3RH\\
$^6$ Department of Astronomy, University of Cape Town, Rondebosch 7700, 
South Africa\\
$^7$ Department of Physics, University of California,
San Diego, CA 92093, USA\\
$^8$ Space Telescope Science Institute, 3700 San Martin Drive, Baltimore, 
MD 21218, USA\\
$^9$ Department of Physics, University of California, Santa Barbara, CA 93106, 
USA\\
$^10$ Departments of Astronomy and Physics, University of Washington, Seattle, 
WA 98195, USA\\
$^{11}$ Department of Physics and Astronomy, Mc~Master University, Hamilton, 
Ontario, Canada, L8S~4M1\\}

\date{Received
      in original form	}

\maketitle

\begin{abstract}

We present optical photometry of the eclipsing supersoft source, 
CAL~87. These observations comprise long term data 
accumulated as a by-product of the MACHO Project, and high speed white light 
photometry of a single eclipse. We ({\it i}) derive an improved ephemeris 
of $T_o = $ HJD $2450111.5144(3) + 0.44267714(6)E$ for the time of minimum 
light, ({\it ii}) 
find the eclipse structure to be stable over a period of $\sim 4$\,years, 
and ({\it iii}) investigate the colour variation as a function of orbital 
phase. The resolution afforded by the high speed nature of the white light 
observations enables us to see new structure in the light curve morphology. 

\end{abstract}

 \begin{keywords}
accretion, accretion discs -- binaries: close -- binaries: 
spectroscopic -- stars: individual: CAL~87 -- Magellanic Clouds -- X-rays:stars
 \end{keywords}

\section{Introduction}

CAL~87 is one of the prototypical members of the new class of low mass X-ray 
binaries (LMXBs) 
which exhibit luminous `supersoft' emission. The supersoft sources 
(SSSs) are characterised by their essential absence of emission above 
$0.5$\,keV, and extremely high luminosities (L$_{\mbox{bol}} \sim 
$10$^{\mbox{38}}$~erg\,s$^{-1}$), of the order of the Eddington limit 
for a 1\,M$_{\odot}$ accretor. 

The original SSSs, CAL~83 and CAL~87, were first detected 
in the Large Magellanic Cloud in 1979-1980 with the 
{\it Einstein} X-ray Observatory (Long, Helfand \& Grabelsky, 1981). 
These systems exhibit the characteristic hallmarks of LMXBs, 
namely blue continua with strong emission lines of HeII~4686 and H$\alpha$
(e.\,g.\ Crampton et al.\ 1987; Pakull et al.\ 1988). 

{\it ROSAT} observations have shown the SSSs to comprise a range of objects, 
including a planetary nebula nucleus (Wang 1991) 
and symbiotic systems (e.\ g.\ Hasinger 1994), although the CAL~83 type 
binaries are distinguished by the highest (near Eddington-limited) 
luminosities. 
Although theories of black hole (Cowley et al.\ 1990) and neutron star 
accretors (Greiner, Hasinger \& Kahabka 1991; Kylafis \& Xilouris 1993)
have been invoked for the latter systems, the most popular model 
involves a white dwarf undergoing stable nuclear surface burning  
(van den Heuvel et al.\ 1992). This has received observational confirmation 
in at least one system, RX~J0513.9-6951, which exhibits a collimated 
bipolar outflow (Crampton et al.\ 1996; Southwell et al.\ 1996). The 
observed jet velocities are of the order of $3800$\,km\,s$^{-1}$, which 
corresponds to the escape velocity of a white dwarf. Hence the nature of 
the accretor is established in at least one SSS (see e.~g.\ Livio 1997). 
The required accretion rates are high 
($ \simgt 10^{-7} {\rm M}_{\odot}$~yr$^{-1}$), and thus require 
a donor star more massive than the white 
dwarf, in order to sustain thermally unstable mass transfer. 

CAL~87 has been optically identified with a $V \sim 19$ mag blue star in 
the LMC (Pakull et al.\ 1988). Optical photometry revealed the system to be 
eclipsing, with an orbital period of 10.6\,h (Callanan et al.\ 1989; 
Schmidtke et al.\ 1993). Although Cowley et al.\ (1990) proposed the 
accretor to be a black hole from an analysis of the He{\sc ii}~4686 
velocities, this identification is highly questionable, particularly since 
accretion disc winds may bias the measurements (van den Heuvel et 
al.\ 1992). Recently, Schandl, Meyer-Hofmeister \& Meyer (1996) 
were able to successfully model the gross features of the optical light curve, 
assuming the model of van den Heuvel et al.\ (1992). 

We present here the results of $\sim 4$~years of optical monitoring of the 
source, the most extended coverage of this system to date. 
These observations were acquired as a by-product of the MACHO project 
(Alcock et al.\ 1995a), owing to the serendipitous location of CAL~87 in a 
surveyed field (see also Alcock et al.\ 1996). Furthermore, we have acquired 
high speed white light photometry of a single eclipse, enabling us to 
investigate the detailed structure. 

\section{Observations and Reduction}

The MACHO observations were made using the 
$1.27$-m telescope at Mount Stromlo Observatory, Australia. 
A dichroic beamsplitter and filters provide simultaneous CCD photometry 
in two passbands, a `red' band ($\sim 6300-7600$~\AA) 
and a `blue' band ($\sim 4500-6300$~\AA). The latter filter is approximately 
equivalent to the Johnson $V$ passband (see Alcock et al.\ 1995a for further 
details). The images were reduced with the standard 
MACHO photometry code {\sc SoDoPHOT}, based on 
point-spread function fitting and 
differential photometry 
relative to bright neighbouring stars. 
Further details of the instrumental set-up and data processing 
may be found in Alcock et al.\ 1995b, Marshall et al.\ 1994 and 
Stubbs et al.\ 1993. 

The high-speed white light observations were made on 28/1/96, using the SAAO 
1.9~m telescope at Sutherland, South Africa. The detector was the UCT 
CCD, operating in frame transfer mode. In this configuration, only half of 
the chip is exposed, and at the end of the integration, the signal is read 
out through the masked half. In this way, we are able to obtain consecutive 
integrations with virtually no CCD dead time. The effective wavelength of 
the instrument, in the absence of a filter, is approximately that of the 
Johnson $V$ passband, but with a bandwidth of $\sim 4000$\AA\ (FWHM). We 
initially used 10~s integrations, but later increased the exposure time to 
20~s, to obtain better signal-to-noise. The images were reduced using the 
profile-fitting technique implemented in DAOPHOT (Stetson 1987).

\section{MACHO Project Photometry}

\subsection{Overall morphology}

We show in Fig.~1 the folded blue and red light curves. Each passband 
comprises 495 observations acquired from 1992~Aug -- 1996~May, 
averaged into 50 phase bins. We use the ephemeris 
$T_o = $ HJD $2450111.5144(3)$ for the time of minimum light (see Sec.~4)
and an orbital period of $P=0.44267714$~d (see Sec.~3.2 below). 
One-sigma error bars are shown and we plot two cycles for clarity. 
The absolute calibration of the MACHO fields 
and transformation to standard passbands is not yet complete, 
thus only the relative magnitudes are plotted. 

We note that the light from CAL\,87 is contaminated by two field stars which 
lie within 
$1''$ away, and have recently been classified through HST observations as 
early-G and late-A type 
stars (Deutsch et al.\ 1996). The seeing at Mount Stromlo Observatory did 
not allow us to resolve these contaminants, thus the MACHO light curves 
include the contributions from these stars. 

The principal features of the light curves are the deep, extended primary
eclipse and shallower secondary dip at $\phi \sim 0.5$ (phase zero is defined 
as inferior conjunction of the donor star). During the primary eclipse, the 
contaminant stars contribute $\simlt 60\%$ of the total light in $V$ and $R$ 
(Deutsch et al.\ 1996). Thus, we do not see the full eclipse depth of 
CAL~87 since, at primary minimum, the light is dominated by the 
contribution of the contaminant stars. For example, in the blue light curve of 
Fig.~1, we see a primary eclipse depth of $\sim 1.2$~mag, compared with an 
expected intrinsic amplitude of 1.57~mag (Deutsch et al.\ 1996). The optical 
observations of Cowley et al.\ (1991) included a deblended $V$ light curve 
with a primary eclipse depth of $\sim 1.4$~mag; this is consistent with 
the fact that one of the contaminant stars is not resolvable on their 
ground-based images, and was thus still contributing to the measured light. 

Despite these contaminants, the light curve {\it structure} is unaffected. 
The extended ingress of 
the primary eclipse is probably due to obscuration by an accretion disc bulge
(e.~g.\ Callanan et al.\ 1989), since the 
secondary itself could not occult the hot inner disc regions at phases 
$\phi \sim 0.7$. 
Following the initial decline, which begins at $\phi \simgt 0.7$, there is a 
`shoulder' in the ingress throughout phases $\phi \sim 0.8-0.9$. Furthermore, 
during the primary eclipse egress, we see marginal evidence for a 
step at $\phi \sim 0.1$. However, these features are far more obvious 
in our high speed light curve (Fig.~5), thus we reserve a discussion for 
Sec.~4.

A secondary dip is observed in both blue and red light curves at 
$\phi \sim 0.5$. In both cases, the amplitude of the dip is around 0.2~mag 
(we see the full intrinsic effect since, at this phase, the light from 
CAL~87 dominates the contaminants). 
A likely cause of the secondary dips is obscuration of 
the irradiated face of the donor star (van den Heuvel et al.\ 1992), 
although Callanan et al.\ (1989) favour an explanation in terms of a 
second accretion disc bulge, which occults the hot, inner disc regions. 

\subsection{Light curve stability and period analysis}

In order to investigate the stability of the eclipse structure, we divided 
the data into four sections, each representing approximately one year's 
worth of data. These red and blue light curves are shown in Figs.~2 and 3.  
We see no significant changes in the overall morphology in either set of light 
curves during $\sim 4$~y from 1992-1996; furthermore, if we consider the 
data of Schmidtke et al.\ (1993) and Callanan et al.\ (1989) the stability 
may be extended back to $\sim 1987$. 
However, we do note that the step in the 
egress at $\phi \sim 0.1$ sometimes seems more pronounced, particularly in the 
blue data of 1993-1994 (see Sec.~4 for an explanation). 

We performed a power spectrum 
analysis on the red and blue datasets.  
A Lomb-Scargle (Lomb 1976; Scargle 1982) analysis indicated an orbital 
period of $P=0.442676 \pm 0.000007$~d, consistent with 
previous measurements (Schmidtke et al.\ 1993). 
However, combining our ephemeris (see Sec.~4) with 
that of Schmidtke et al.\ (1993) allows us to derive an improved orbital 
period of $P=0.44267714 \pm 0.00000006$~d. 
A period analysis of each dataset failed to reveal any significant  
change in the orbital period over the four years of monitoring. From a 
consideration of the accuracy with which we are able to determine the period 
from four years of data, we derive an upper limit for the magnitude of any 
orbital period changes of $\dot{P} \simlt 7 \times 10^{-8}$. This may be 
used to obtain an approximate upper limit on the mass transfer rate
(e.~g.\ Robinson et al.\ 1991), by 
assuming the main sequence mass-radius relation for the secondary 
(e.~g.\ Echevarr\'{\i}a 1983), an 
equivalent volume Roche lobe radius relation (Paczy\'nski 1971) and Kepler's 
law. However, we find that the constraint on 
$\dot{P}$ is not stringent enough to yield a particularly useful result. We 
find an approximate upper limit on $\dot{M}$ of $\sim 
7 \times 10^{-5}~M_{\odot}$~yr$^{-1}$, which merely 
indicates that the accretion rate is sub-Eddington.

\subsection{Orbital colour variation}

As discussed in Sec.~3.1, the MACHO light curves represent the blended light 
of CAL\,87 and the contaminant stars. 
However, the spectral types of these stars are 
sufficiently early that they have a virtually negligible effect 
on the orbital $V-R$ colour variation of CAL\,87. 

To demonstrate this, we plot the relative $V-R$ colour, folded on the orbital 
period, as a function of orbital phase in Fig.~4. 
We see a pronounced reddening of $\sim 0.16$~mags during the primary eclipse. 
However, we can be confident that this is not simply due to the changing 
contribution of the contaminant stars, since Deutsch et al.\ (1996) have 
quantified the effect. They observed the colour change during primary eclipse 
for both intrinsic and blended light curves, finding that the 
amplitude differed by only $\sim 0.03$~mags between the two cases. 
Therefore, since we see a much larger effect, we conclude that this is an 
intrinsic variation which is consistent with the occultation of a hot, blue 
inner disc. Our value may be compared with the change 
found by Deutsch et al.\ (1996) of 0.21~mag in the blended $V-R$ colour. 

There is also marginal evidence for 
a slight reddening centred on $\phi \sim 0.55$. This may be caused by the 
partial obscuration of the inner disc or the heated face of the donor star, 
although it is not clear whether the phase delay of 0.05, compared with the 
time of secondary minima in Fig.~1, is significant. 
We note that 
previous optical $V$-band observations (Callanan et al.\ 1989; Cowley et al.\ 
1990; 
Schmidtke et al.\ 1993) indicate that the phase of the secondary minima 
in the optical light curves may be slightly variable ($\phi \sim 0.45-0.55$). 
The reddening throughout the primary eclipse is $\simgt 5$ times greater than 
in the dip at $\phi \sim 0.55$, although we stress that the latter may not be 
significant. Indeed, we find no explanation for the apparently anomalous 
data points at $\phi \sim 0.32$ and $\phi \sim 0.12$. 

\section{High speed photometry}

We show in Fig.~5 the `real time' white light curve obtained from the SAAO 
observations. We plot the white light counts relative 
to a local standard, although the absence of a filter prevents us from 
making an absolute calibration (the light is also blended with the 
contaminant stars). 
The errors on the data points are 
$\sim 0.005$ counts, but rise for $\phi \simgt 0.15$ (these observations were 
made during morning twilight). 
We note a decrease in the scatter of the light curve for $\phi \simgt 
0.85$, when we switched from 10~s to 20~s integrations. However, it is 
also possible that we are seeing a visible contribution from the diminished 
effects of flickering as the disc is eclipsed, a 
phenomenon commonly observed in cataclysmic variables (see e.~g.\ Warner 1995). 
The time of minimum light was derived by fitting a parabola to the most 
symmetrical part of the eclipse centre ($\phi \sim 0.95-0.05$), and is 
found to be $T_o = $ HJD $2450111.5144(3) + 0.44267714(6)E$. 

The detailed morphology of the eclipse structure reveals several interesting 
features. In Fig.~5, we see an obvious shoulder in the ingress at 
$\phi \sim 0.82-0.87$, which is also apparent in the MACHO light curves of 
Fig.~1 (see also Callanan et al. 1989; Cowley at el. 1990). 
The origin of this feature may probably 
be understood in terms of the variable obscuration by the 
thickened accretion disc rim. Recently, Armitage \& Livio (1996) have 
performed three-dimensional smooth particle hydrodynamic (SPH) 
simulations of accretion discs in LMXBs. In particular, they compute the 
column density profile towards the primary for a system 
inclination of $78^{\circ}$, 
which is probably appropriate for CAL~87 (Schandl, Meyer-Hofmeister \& Meyer 
1996). They find 
a strong peak centred on $\phi \sim 0.8$, which has declined  
by $\sim 40\%$ at $\phi \sim 0.9$. Thus, the gradient of our light 
curve ingress appears to be 
qualitatively consistent with these model calculations, being steepest at the 
phase of maximum obscuration ($\phi \sim 0.8$), but levelling off as the 
column density declines towards $\phi \sim 0.9$. 
The rapid decline in the luminosity for $\phi > 0.9$ presumably occurs as the 
donor star itself begins to obscure our line of sight. 

Within the deepest part of the primary eclipse ($\phi \sim 0.9-0.1$), we see 
suggestions of further structure at $\phi \sim 0.92, 0.96, 0.02$ and $0.06$. 
However, it is not clear that we can attach any physical significance to 
these points, given the accuracy of the data, and further 
observations are scheduled in order to investigate the repeatability of these 
features. Nevertheless, 
a very clear variation is observed in the egress for $\phi \simgt 0.1$. In 
this respect, the light curve of CAL~87 is reminiscent of certain 
eclipsing cataclysmic variables (e.~g.\ Z~Cha; Cook \& Warner 1984). In the 
latter system, a marked turn-over is observed in the eclipse egress after 
the emergence of the primary star; then, following a 
short delay, the light curve rises again. 
This behaviour is attributed to the 
emergence of a `hot spot,' a highly luminous region on the disc rim caused by 
the accretion stream/disc interaction. In CAL~87, an analogous component may 
be the strongly irradiated inner-surface of the accretion disc bulge. Indeed, 
Schandl, Meyer-Hofmeister and Meyer (1996) note that the existence of such a 
component is consistent with the 
fact that the luminosity is greater following an eclipse. The MACHO light 
curves of Fig.~2 and Fig.~3 suggest that this feature is at some times more 
pronounced than others (see Sec.~3.2). 
This is probably not 
surprising, if indeed we are seeing the emergence of a variably irradiated 
region. 

\section{Conclusions}

We find the light curve of CAL~87 to be essentially stable over a four 
year period from 1992-1996, with an overall morphology consistent with earlier  
observations. 
We resolve structure in the primary eclipse at $\phi \sim 0.8-0.9$ 
and $\phi \sim 0.1$. These features are consistent with the presence of an 
accretion disc bulge, with a bright, irradiated inner surface. 
The orbital $V-R$ colour shows a pronounced reddening at 
primary eclipse and a probable smaller effect at $\phi \sim 0.55$, 
due to the obscuration 
of the primary/inner disc regions and the irradiated donor star respectively. 
We derive an improved orbital period of $P=0.44267714(6)$~d, finding 
no evidence for a significant change in $\simgt 4$ years.

\subsection*{Acknowledgments}

We are grateful for the support given our project by the technical
staff at the Mt. Stromlo Observatory. Work performed at LLNL is 
supported by the DOE under contract W-7405-ENG. Work performed by the
Center for Particle Astrophysics personnel is supported by the NSF 
through grant AST 9120005. The work at MSSSO is supported by the Australian
Department of Industry, Science and Technology. 
KG acknowledges support from DoE OJI, Alfred P. Sloan, and Cotrell Scholar 
awards. 
CWS acknowledges the generous support of the Packard and Sloan Foundations.
WS and KAS are both supported by PPARC through an Advanced Fellowship and 
studentship respectively. KAS thanks Paul Callanan for a useful 
discussion, and F.\ van Wyk for his technical support.

\clearpage
\newpage

\begin{figure}
\caption{The blue ({\it upper}) and red ({\it lower}) optical light curves of 
CAL~87, blended with the contaminant stars (see Sec.~3.1), 
acquired from MACHO Project observations during 1992-1996. The data are 
folded on $T_o =$ HJD $2450111.5144(3) + 0.442674(7)E$ 
and averaged into 50 phase bins. Two cycles are plotted for clarity. The 
primary eclipse depth is affected by the dominating light of the field stars.} 
\end{figure}

\begin{figure}
\caption{Blue light curves of CAL~87 + contaminants, split into $\sim 1$~y 
intervals. 
Two cycles of the unbinned folded data are plotted.
The light curve morphology is essentially stable during the four years of 
observations.}
\end{figure}

\begin{figure} 
\caption{Red light curves of CAL~87 + contaminants, split into $\sim 1$~y 
intervals. 
Two cycles of the unbinned folded data are plotted.
The light curve morphology is essentially stable during the four years of 
observations.}
\end{figure}
 
\begin{figure}
\caption{Relative $V-R$ colour of CAL~87 (and contaminant stars) plotted as a 
function of orbital phase. Note the pronounced reddening during the primary 
eclipse, and possible smaller dip at $\phi \sim 0.55$.} 
\end{figure} 

\begin{figure}
\caption{SAAO high speed white light curve of CAL~87 + contaminants. The data 
are not folded 
and consist of 10~s and 20~s integrations. Typical errors are $\sim 0.005$ 
counts. Note the obvious structure at 
$\phi \sim 0.85$ and $\phi \sim 0.1$ (see Sec.~4).}
\end{figure} 

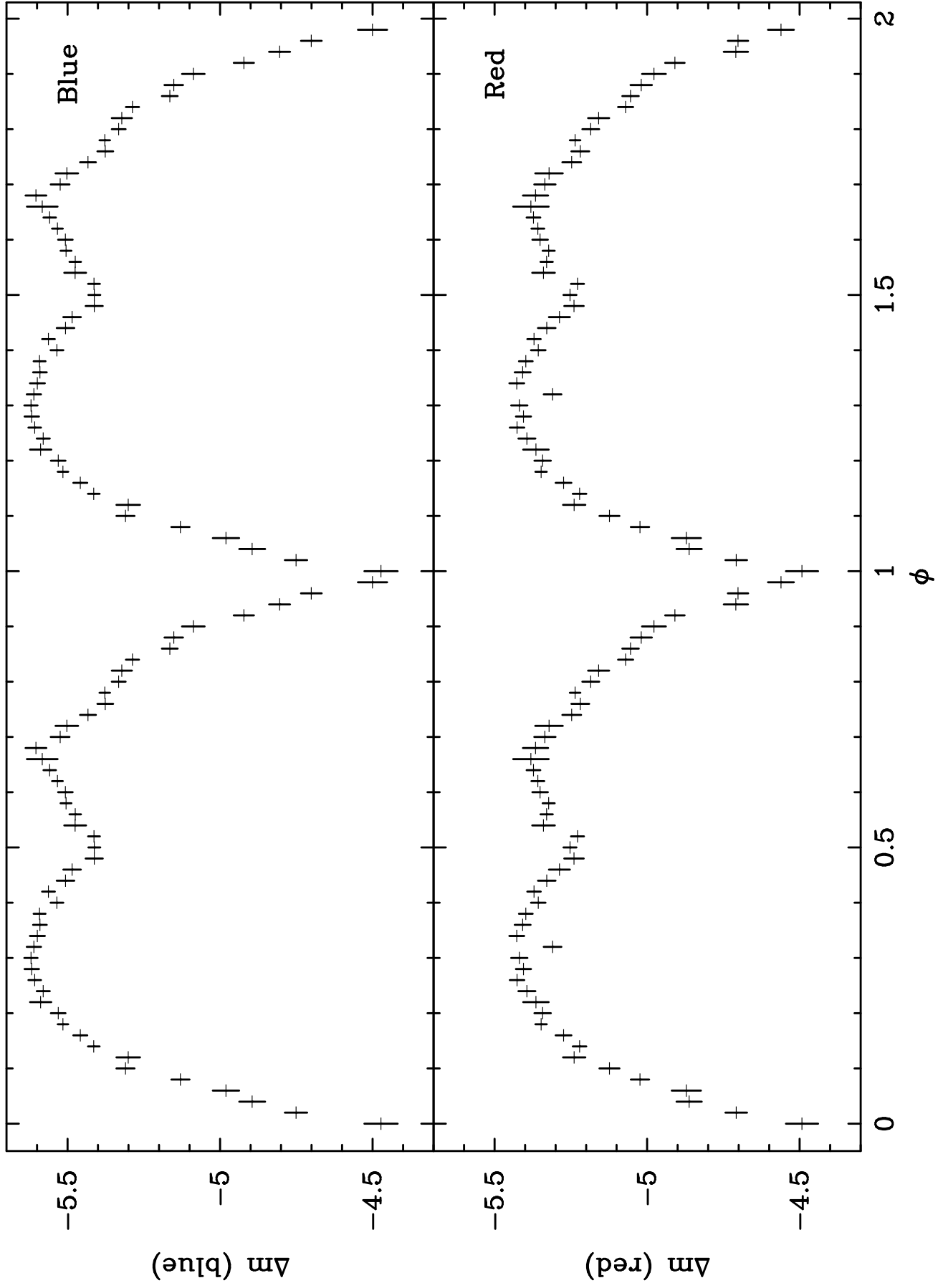
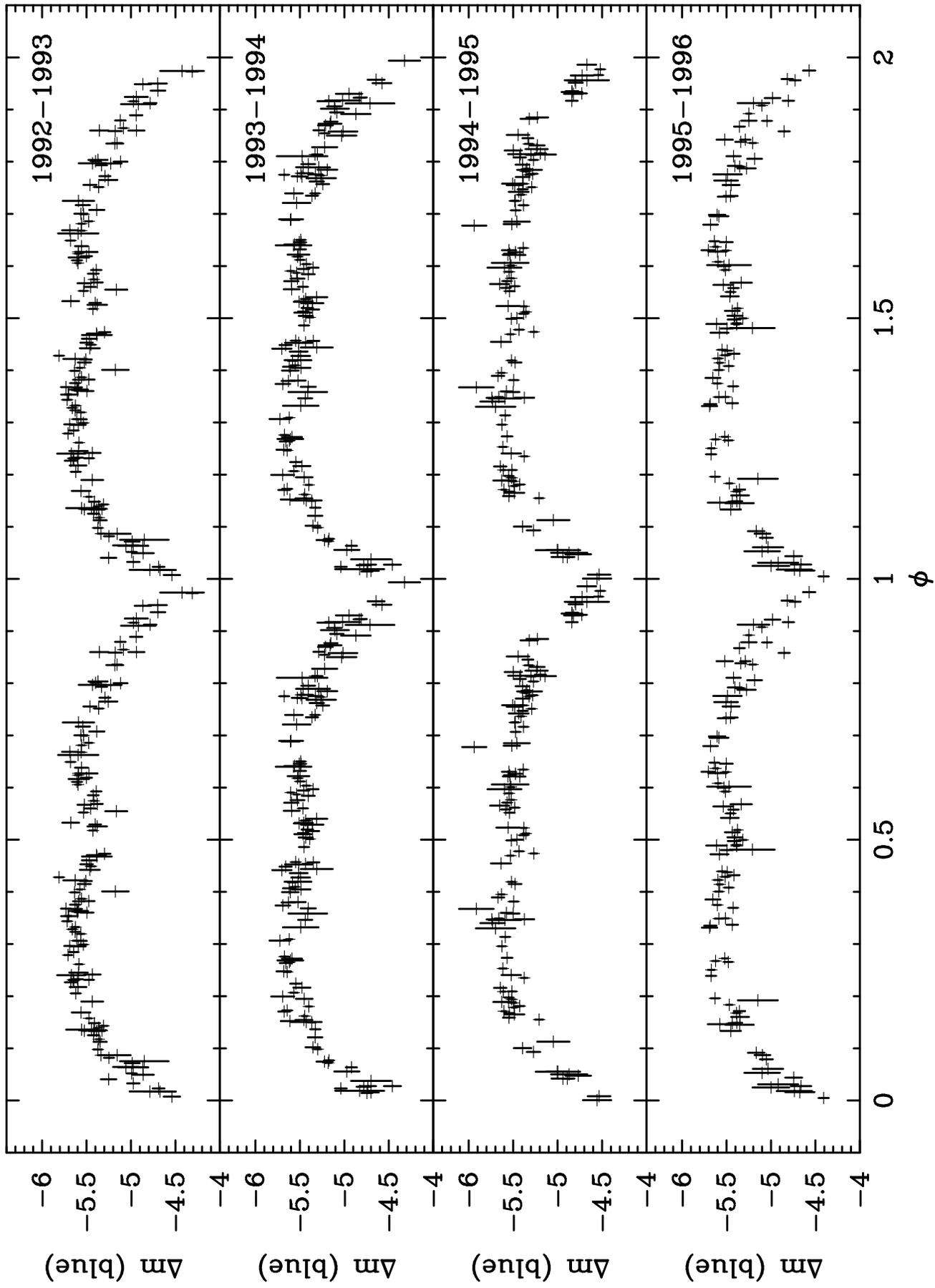
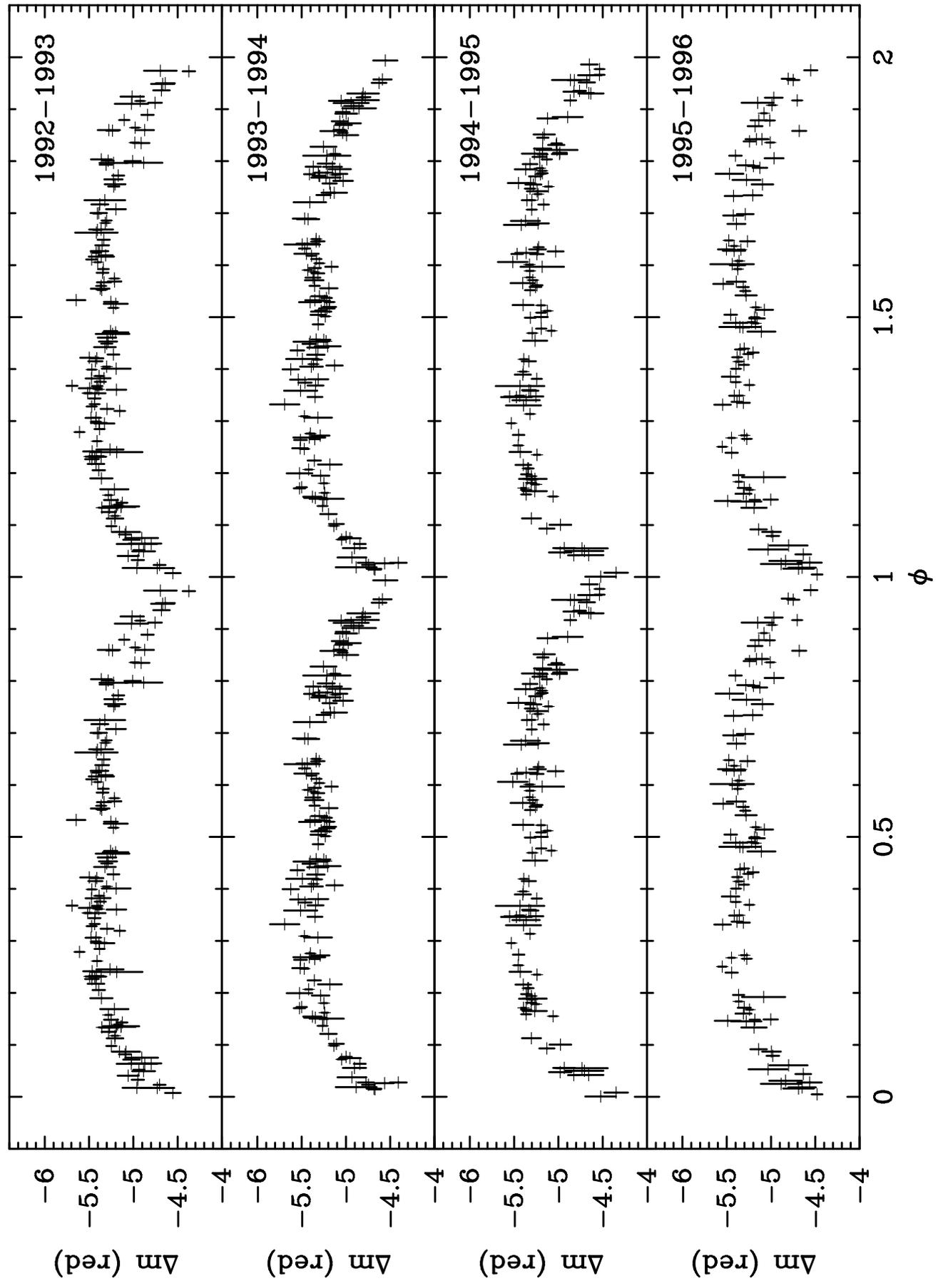
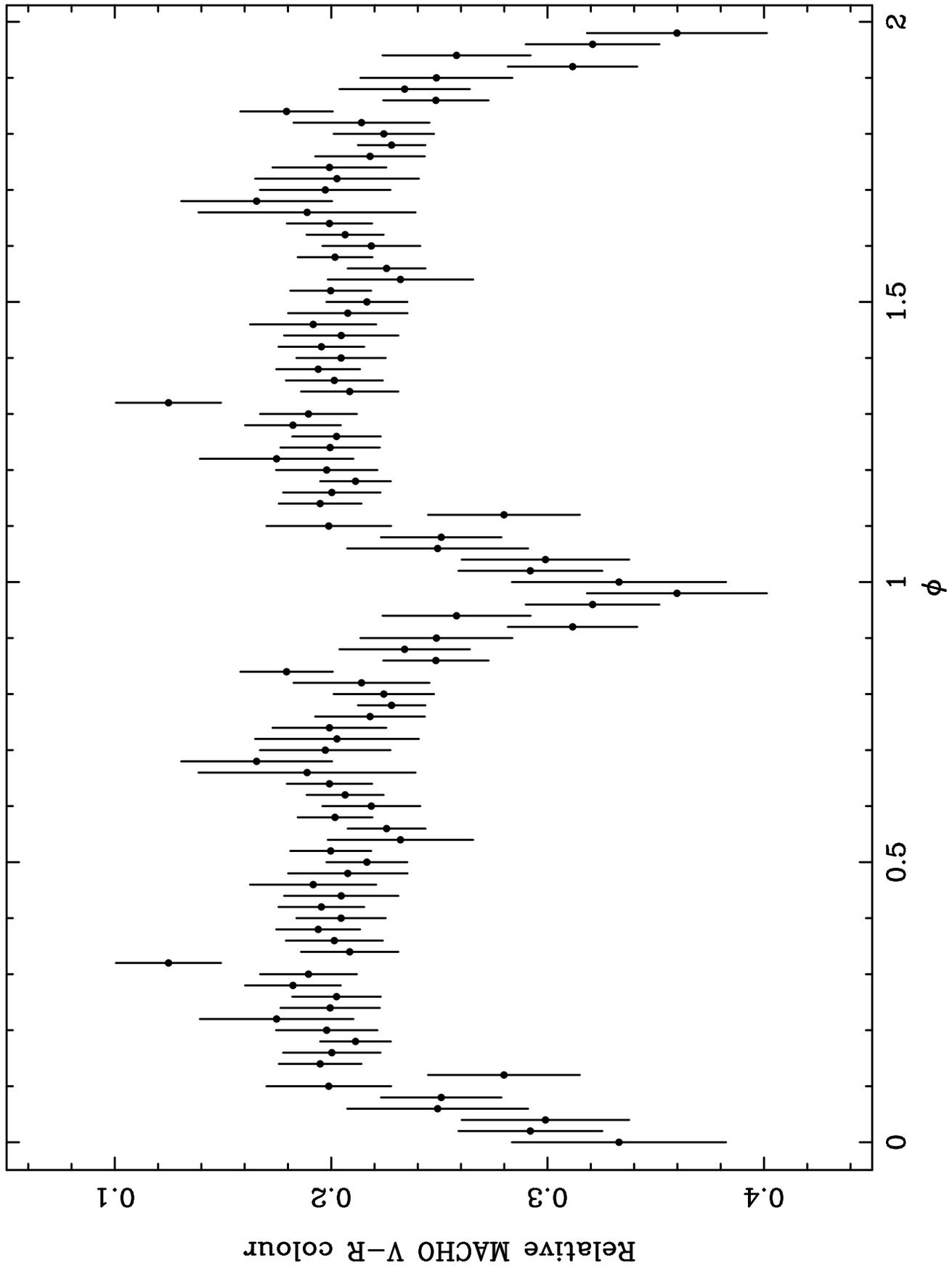
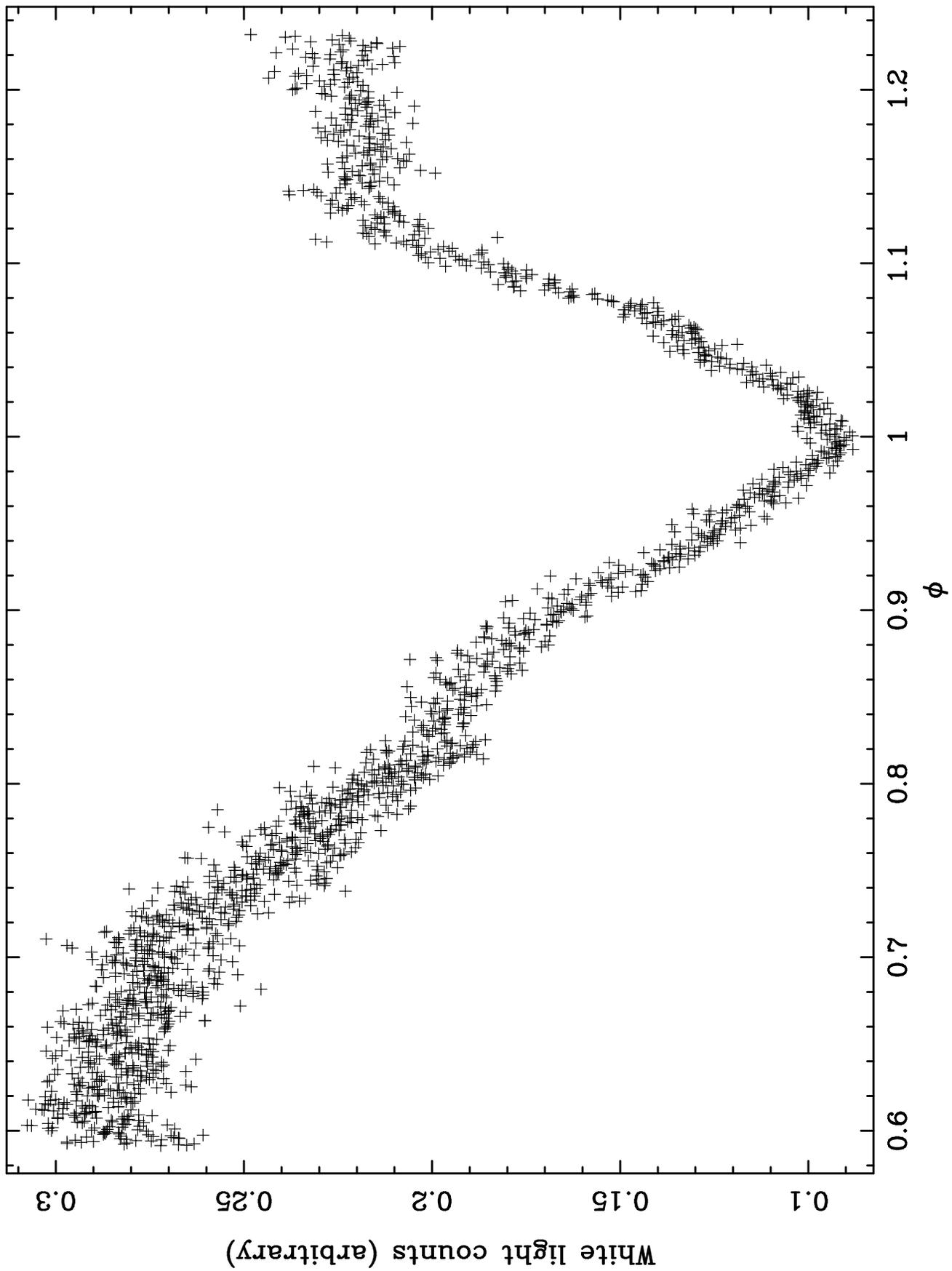
\end{document}